\documentclass[aps, pre, floatfix, superscriptaddress, nofootinbib]{revtex4}
\usepackage{mathrsfs} \usepackage{graphicx,color} \usepackage{bbm} \usepackage{multirow}

\usepackage{amsmath} 
\usepackage{amssymb}
\usepackage{amsfonts}
\usepackage{latexsym}
\usepackage{amsthm}
\usepackage{varioref}
\usepackage{listings}
\usepackage{float}
\usepackage{rotate}
\usepackage{epsfig}
\usepackage{psfrag}
\usepackage{mathtools}
\usepackage{longtable}

\usepackage{graphicx, color}

\definecolor{darkred}{rgb}{0.9,0,0.3}
\definecolor{darkblue}{rgb}{0,0.3,0.9}

\let\G=\Gamma

\def\Re{{\rm Re}\,}\def\Im{{\rm Im}\,}

\def\BBB{{\bf W}}

\def\Xint#1{\mathchoice
   {\XXint\displaystyle\textstyle{#1}}%
   {\XXint\textstyle\scriptstyle{#1}}%
   {\XXint\scriptstyle\scriptscriptstyle{#1}}%
   {\XXint\scriptscriptstyle\scriptscriptstyle{#1}}%
   \!\int}
\def\XXint#1#2#3{{\setbox0=\hbox{$#1{#2#3}{\int}$}
     \vcenter{\hbox{$#2#3$}}\kern-.5\wd0}}

\def\dashint{\Xint-}

\numberwithin{equation}{section}

\renewcommand{\b}[1]{\boldsymbol{\mathrm{#1}}} 
\renewcommand{\cal}{\mathcal}

\newcommand \M {\textbf{M}}
\newcommand \C {\textbf{C}}

\newcommand \mso{\Phi}

\newcommand \be  {\begin{equation} }
\newcommand \ee  {\end{equation}}
\newcommand \Tr {\text{Tr}}

\renewcommand{\G}{\textbf{G}}
\newcommand{\stj}{g}

\renewcommand \l{\lambda}

\newcommand{\ii}{\mathrm{i}}
\newcommand{\dd}{\mathrm{d}}

\newcommand*{\deq}{\mathrel{\vcenter{\baselineskip0.65ex \lineskiplimit0pt \hbox{.}\hbox{.}}}=}

\renewcommand{\leq}{\leqslant}

\renewcommand{\geq}{\geqslant}

\renewcommand{\epsilon}{\varepsilon}

\newcommand{\pB}[1]{\Bigl({#1}\Bigr)}

\newcommand{\pBB}[1]{\Biggl({#1}\Biggr)}

\newcommand{\qb}[1]{\bigl[{#1}\bigr]}

\newcommand{\scalar}[2]{\langle{#1} \mspace{2mu}, {#2}\rangle}

\newcommand \beno  {\begin{equation*}}
\newcommand \bea {\begin{eqnarray} \nonumber }
\newcommand \eeno  {\end{equation*}}
\newcommand \eea {\end{eqnarray}}

\begin{document}

\title{Two short pieces around the Wigner problem}
\author{Jean-Philippe Bouchaud$^1$, Marc Potters}
\affiliation{Capital Fund Management, 23--25, rue de l'Universit\'e, 75\,007 Paris}
\begin{abstract}
We revisit the classic Wigner semi-circle from two different angles. One consists in studying the Stieltjes transform directly on the real axis, which
does not converge to a fixed value but follows a Cauchy distribution that depends on the local eigenvalue density.
This result was recently proven by Aizenman \& Warzel for a wide class of eigenvalue distributions. We shed new light onto their result using a Coulomb gas method. The second angle is to derive a Langevin equation for the full (matrix) resolvent, extending Dyson's Brownian motion framework. The full matrix structure of this equation allows one to recover known results on the overlaps between the eigenvectors of a fixed matrix and its noisy counterpart.
\end{abstract}
\maketitle

\section{Introduction}

Wigner's semi-circle law is certainly the most famous in Random Matrix Theory. There are many different ways to obtain it, each of them shedding a different light on the result. 
In the present paper written for this special issue on Random Matrix Theory, we revisit once again the Wigner problem. In the first part of this paper, we show that the semi-circle can be obtained by studying the (power-law) tail of the normalized trace of the resolvent $g(x)$ {\it on the real axis}, rather than ``just above'' the real axis in the complex plane. 
On the support on the eigenvalue density, $g(x)$ does not converge to a constant value (the limiting Stieltjes transform $g_0(x)$ is ill-defined for such $x$) but rather converges in probability to a Cauchy law, a result obtained by Fyodorov and collaborators \cite{Fyodorov1, Fyodorov2, Fyodorov3} for the GOE and the GUE,\footnote{Note that this result had fact previously appeared in  P. A. Mello, ''Mesoscopic Quantum Physics,'' Les Houches Summer School, Edts. E. Akkermans et al., Elsevier, Amsterdam, 1995, Session LXI, p. 435} and later proven by Aizenman \& Warzel \cite{Aizenman} for a very large class of point processes. We re-derive this result using a Coulomb gas method, which amounts to study the response of such a gas to a singular perturbation. The second part of our paper follows Dyson's Brownian motion framework to derive a Langevin equation for the full (matrix) resolvent. This Langevin equation becomes deterministic in the large $N$ limit; its trace leads to the well known Burgers equation the solution of which again produces Wigner's semi-circle. But the full matrix structure allows one to characterize the evolution of the eigenvectors as well, and recover known results on the overlaps between the eigenvectors of a fixed matrix $\C$ and its noisy counterpart $\C + \BBB$, where $\BBB$ is the Wigner-Dyson Brownian random matrix. The case of an isolated (spike) eigenvalue is also discussed within the same framework.    

\section{The resolvent of a Wigner matrix on the real axis is Cauchy distributed}

\subsection{Schur elimination and the Cauchy fixed point}
\label{2.1}
A standard way to approach the distribution of eigenvalues of random matrices is to write a recursion relation for the elements of the resolvent matrix $\bf G$, defined as
\be
{\bf G}(z) = (z\mathbb{I} - {\bf W})^{-1},
\ee
where $z$ is in the complex plane, but outside the real axis to avoid the poles of $\bf G$ (i.e. the eigenvalues of ${\bf W}$). The standard Schur relation then allows one to relate the elements of the resolvent matrix for a problem of size $N$ and the same problem with one row and one column added to the matrix 
$\bf W$. Denoting conventionally be ``0'' the index of the added row and column, one readily finds:
\be\label{eq-iter}
\frac{1}{G_{00}^{(N+1)}} = z - W_{00} - \sum_{i,j=1}^N W_{0i} G_{ij}^{(N)} W_{j0}.
\ee
For Wigner random matrices with independent entries of order $N^{-1/2}$, one can further argue that the contributions of terms with $i \neq j$ in the above formula are 
negligible in the large $N$ limit, leading to:
\be\label{iteration}
\frac{1}{G_{00}^{(N+1)}} \approx z - \sum_{i}^N W_{0i}^2 G_{ii}^{(N)}.
	\ee
A crucial point, which makes this formula useful, is that the new matrix elements $W_{0i}$ and the resolvent elements $G_{ii}^{(N)}$ are completely independent. 

As recalled above, the standard route is to study the above iteration for $z$ outside the real axis, in which case the diagonal elements of $\bf G$ converge, for large $N$, to the normalized trace (or Stieltjes transform)
\be
{g_0}(z) := \lim_{N\to\infty} \frac1N \text{Tr}\,{\bf G}(z),
\ee
where $g_0(z)$, is the solution of
\be \label{gz}
\frac{1}{{g}(z)} = z - \sigma^2 {g}(z),
\ee
where $\sigma^2:=N\mathbb{V}[W_{0i}]$. One then recovers the classic result, from which the semi-circle law ensues:
\be
{g_0}(z) = \frac{1}{2 \sigma^2} \left[z \pm \sqrt{z^2 - 4 \sigma^2}\right] \to \rho(x) = \frac{1}{\pi} \lim_{\varepsilon \to 0} \Im {g_0}(x - \text{i}\varepsilon) = \frac{1}{2 \pi \sigma^2}
\sqrt{4 \sigma^2 - x^2}.
\ee
For $z=x$ real and within Wigner's band $[-2 \sigma, 2 \sigma]$, ${g}(x)$ cannot be well defined, because by definition $x$ is then always very close to a pole of ${\bf G}$. An idea, proposed in the context of L\'evy matrices in \cite{cizeau}, is to turn this predicament on its head and actually exploit the divergence of ${g}(x)$ when $x$ is equal to any eigenvalue of ${\bf W}$. In a hand-waving manner, the 
probability that the difference $d_i=|x - \lambda_i|$, between $x$ and a given eigenvalue $\lambda_i$, is very small is: 
\be
\mathbb{P}[d_i < \epsilon/N] = 2 \epsilon \rho(x),
\ee
where $\rho(x)$ is the normalized density of eigenvalues around $x$. But as $\epsilon \to 0$, the resolvent becomes dominated by a unique contribution -- that of the $\lambda_i$ term. In other words,
${g}(x) \approx \pm (N d_i)^{-1}$, and therefore
\be
\mathbb{P}[|{g}| > \epsilon^{-1}] = \mathbb{P}[d_i < \epsilon/N] = 2 \epsilon \rho(x).
\ee
Hence, the tail of the distribution of non-self averaging Stieltjes transform ${g}$ should decay precisely as $\rho(x)/{g}^2$. Studying this tail allows one to extract the eigenvalue density $\rho(x)$ while working directly on the real axis. This was the strategy used used to \cite{cizeau} to obtain the eigenvalue density of L\'evy matrices (see below), a result later revisited in \cite{Burda} and rigorously established by Ben Arous and Guionnet in \cite{BAG}. Here, we want to revisit this issue in the standard Wigner case, in order to shed light on an admittedly weird strategy that Ben Arous \& Guionnet ``{\it unfortunately [could not] make sense of}'' \cite{BAG}.    
\begin{figure} 
\begin{center}
\includegraphics[scale=0.8]{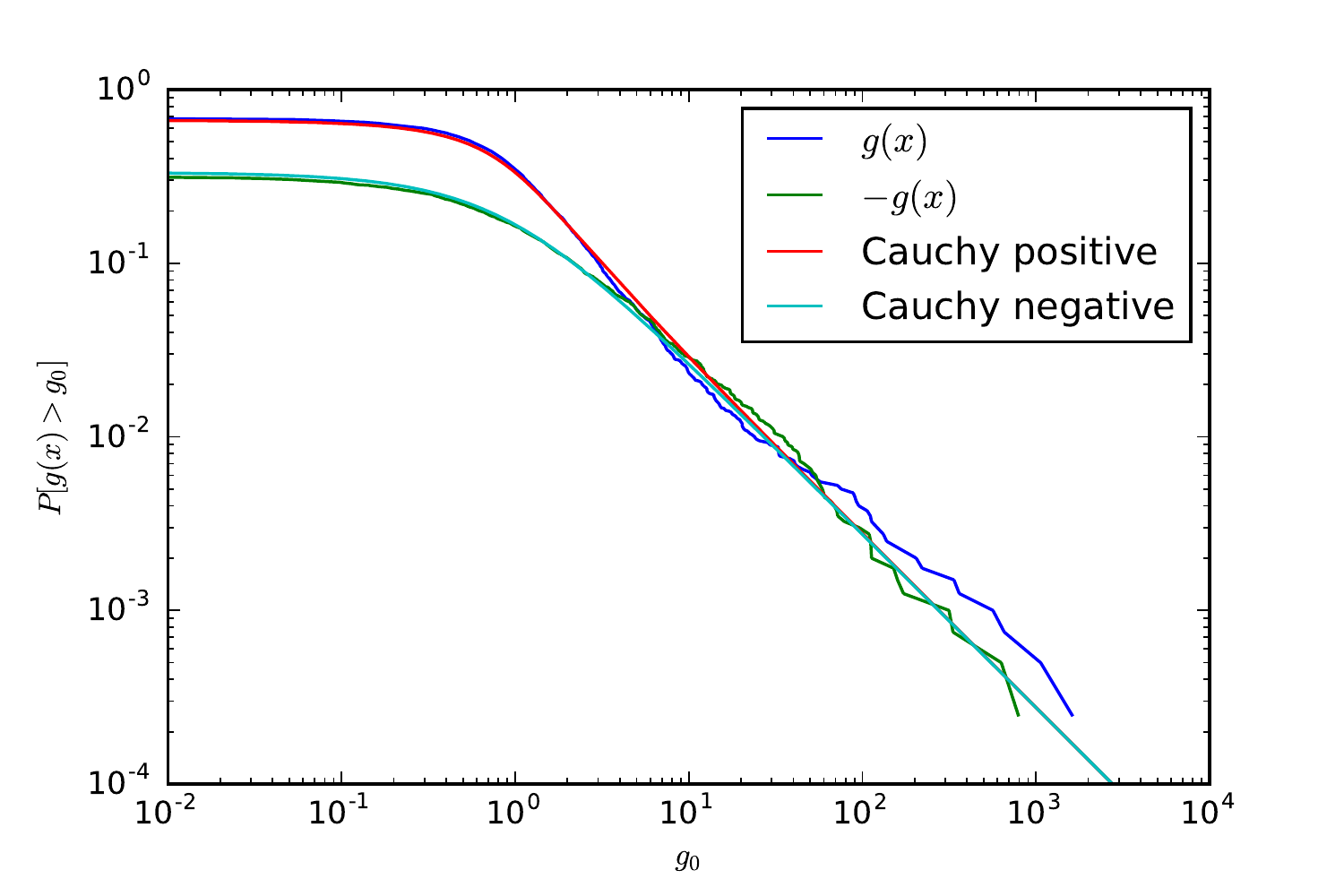}
\caption{Numerical simulation of the law of $g(x)$ for a GOE matrix. For a
fixed $N=5000$ GOE matrix, we have sampled 4000 times the distribution
of $g(x+u/\sqrt{N})$, with $x=1$ and $u$ distributed uniformly between -1 and 1.
Plotted is the left and right sample cumulative probability with the
analytical result for the corresponding Cauchy distribution.}
\end{center}
\end{figure}

The first remark is that for a rotationally invariant problem, the distribution of a randomly chosen diagonal element of the resolvent (say $G_{00}$) is the same as the distribution $P({g})$ of its normalized trace. Therefore, Eq. \ref{iteration} can be interpreted as giving the evolution of $P({g})$ itself, i.e.:
\be
P^{(N+1)}({g}) = \int_{-\infty}^{+\infty} {\rm d}{g}' P^{(N)}({g}') \delta\left({g} - \frac{1}{x - \sigma^2 {g}'}\right),
\ee
where we have used the fact that for large $N$, $\sum_{i}^N W_{0i}^2 G_{ii}^{(N)} \to \sigma^2 {g}^{(N)}$. Now, this functional iteration admits the following Cauchy distribution 
as a fixed point \cite{Hakim}:\footnote{In fact, the presence of a small noise, indeed contained in Eq. (\ref{iteration}), allows this fixed point to be stable, see \cite{Hakim}.} 
\be
P^{\infty}({g}) = \frac{\rho(x)}{({g} - \frac{x}{2 \sigma^2})^2 + \pi^2 \rho(x)}.
\ee
This simple result, that the resolvent of a Wigner matrix on the real axis is a Cauchy variable, calls for several comments. First, one finds that $P^{\infty}({g})$ indeed behaves as $\rho(x)/g^2$ for large $g$, as expected. Second, it would have been entirely natural to find a Cauchy distribution for $g$ had the eigenvalues been independent. Indeed, 
since ${g}$ is then the sum of $N$ random variables (i.e. the $1/d_i$'s) distributed with an inverse square power, the generalized CLT predicts that the resulting sum is Cauchy distributed. In the present case, however, the eigenvalues are strongly correlated -- in fact the spectrum is so rigid that the approximation $\lambda_i - \lambda_j \approx (i - j)/(N\rho(x))$ holds locally to a good approximation. In this case, the Cauchy distribution for ${g}$ was actually derived by Y. Fyodorov and collaborators for the GOE and GUE ensembles, using rather specific Random Matrix Theory techniques \cite{Fyodorov1, Fyodorov2, Fyodorov3}. It was recently proven by Aizenman \& Warzel \cite{Aizenman} that the Cauchy distribution is in fact {\it super-universal} and holds not only for all Coulomb gas models, for arbitrary values of $\beta$, but in fact for a much wider class of point processes on the real axis. The gist of the argument of Aizenman \& Warzel is summarized in Appendix A. In the next subsection, we want to give a physicist' approach to the problem for Coulomb gas models, where we recover the super-universal Cauchy distribution, that is also valid for an arbitrary confining potential. We believe that the direct calculation, using a saddle point method, is quite interesting in its own right and could be used to obtained some refined results at finite $N$.

\subsection{From Coulomb to Cauchy}

Let us consider the resolvent of a $\beta$-ensemble matrix $\bf W$ on the real axis:
\be
{g}(x) = \frac1N \sum_{i=1}^N \frac{1}{x - \lambda_i}
\ee
where $\lambda_i$ are the eigenvalues of $\bf W$. The joint distribution of the $\lambda_i$'s is well known to be:\be
{\cal P}(\{\lambda_i\}) = Z \prod_{i < j} |\lambda_i - \lambda_j |^\beta \exp\left[-\frac{N \beta}{2} \sum_i V(\lambda_i)\right]
\ee
The case $\beta=0$ corresponds to independent (Poisson) random eigenvalues for which, as mentioned above, the Cauchy result is a trivial consequence of the generalized CLT. The case $\beta = \infty$ corresponds 
to a perfectly periodic ``crystal'' of eigenvalues, for which an explicit calculation is also possible, see Appendix B.

The potential $V(x)$ can be any confining potential. To simplify the discussion we start by considering the GxE potential\footnote{Note that we set $\sigma^2=1$, so that the spectrum of $\BBB$ is $[-2,2]$.} $V(x)=x^2/2$
and introduce  later an arbitrary potential.

Let us fix an arbitrary value of $x$ within the spectrum $[-2,2]$ and study the characteristic function of the distribution of ${g}$:
\be \label{charac}
\widehat P(k) = \int \prod_i {\rm d}\lambda_i {\cal P}(\{\lambda_i\}) e^{\text{i}\frac{k}{N} \sum_{i=1}^N \frac{1}{x - \lambda_i}}
\ee
Introducing a density field $\rho(\l)$ and neglecting the entropy term, one finds (see \cite{Majumdar} for a detailed account):
\be
\widehat P(k) = Z \int {\cal D}\rho e^{N^2 \left[ \frac{\beta}{2} \int {\rm d}\lambda'{\rm d}\lambda'' \rho(\lambda')\rho(\lambda'') \log |\lambda'-\lambda''| - \frac{\beta}{4} \int {\rm d}\lambda' \rho(\lambda') \lambda'^2\right] 
+ \text{i}k \int {\rm d}\lambda' \frac{\rho(\lambda')}{x - \lambda'}}.
\ee
The saddle point equation on $\rho$ then reads:
\be
\int {\rm d}\lambda' \rho(\lambda') \log |\lambda-\lambda'| - \frac14 \lambda^2 + \text{i}\widehat k \frac{1}{x - \lambda} + K = 0
\ee
where $K$ is the Lagrange multiplier insuring that $\rho(x)$ is normalized, and $\widehat k = N^{-2} k/\beta$. Taking the derivative of this equation w.r.t. $\lambda$ yields:
\be \label{saddle_cont}
\dashint {\rm d}\lambda' \frac{\rho(\lambda')}{\lambda-\lambda'}  = \frac12 \lambda - \text{i}\widehat k \frac{1}{(x - \lambda)^2} 
\ee
For $k=0$, the solution is the familiar Wigner distribution:
\be
\rho_0(\lambda) = \frac{1}{2 \pi} \sqrt{4 - \lambda^2}
\ee
Since the equation for $\rho$ is linear, the solution for $k \neq 0$ simply reads:
\be
\rho(\lambda) = \rho_0(\lambda) + \delta \rho(\lambda); \qquad \delta \rho(\lambda) =  -\text{i}\widehat k \partial_x \delta(\lambda-x).
\ee
This corresponds to a shift of the unperturbed eigenvalues $\lambda_0$ by a {\it singular} quantity $\delta \lambda =  \text{i}\widehat k \delta(\lambda-x)/\rho_0(x)$. So, although this solution is
valid in the continuum limit, we cannot use it directly to estimate ${g}(x)$ for finite $N$. We need to ``zoom'' into the Dirac delta function to resolve the shift on atomic distances. Nevertheless, we will later make use of the above result to fix the asymptotic behaviour of the shifted Stieltjes transform:
\be\label{asympt}
\delta g(z) := \dashint {\rm d}\lambda \frac{\delta \rho(\lambda)}{z-\lambda} \approx  -\text{i}\widehat k \frac{1}{(x - z)^2}, 
\ee
when $|x-z|$ is large enough.

In order to make progress, we write the discrete analogue of the saddle-point equation Eq. (\ref{saddle_cont}) as
\be
\frac1N\sum_{j \neq i} \frac{1}{\lambda_i - \lambda_j} - \frac12 \lambda_i = -\text{i}\widehat k \frac{1}{(x - \lambda_i)^2}
\ee
Multiplying both sides by $(N(z - \lambda_i))^{-1}$ and summing over $i$ leads, after a few standard
manipulations, to an equation for the Stieltjes transform $g(z)$ that reads
(neglecting a $1/N$ contribution):
\be \label{saddle_disc}
 g^2(z) - V'(z) g(z) + P(z) =  -2\text{i}\widehat k \partial_x \left[ \frac{g(z)-g(x)}{z-x} \right],
\ee
where we have introduced a more general confining potential $V(z)$ and associated function
$P(z)$\footnote{$P(z):=\sum_i (V'(z)-V'(\lambda_i))/(z-\lambda_i)$, it is a polynomial 
of degree $n-1$ when $V(z))$ is a polynomial of degree $n$,
see e.g. \cite{PhysRep2}}. For the standard case, $V(z)=z^2/2$ and $P(z)=1$.

Now, the strange thing that happens is that when $z$ is the vicinity of $x$, the right hand side remains of order unity in the limit $\widehat k \to 0$. More precisely, the 
eigenvalue density perturbation turns out to take the following scaling form:
\be\label{eq-defF}
\delta \rho(\lambda) = \text{i} \zeta F\left(\frac{x - \lambda}{\sqrt{|\widehat k|}}\right), \qquad (\widehat k \to 0),
\ee
where $\zeta= \text{sign}(k)$ and $F(u)$ is a certain odd function of $u$. This means that the extra (imaginary) ``charge'' located at 
$\lambda=x$ substantially affects the Coulomb gas density in a neighbourhood of size $\sqrt{|\widehat k|} \sim N^{-1}$ around $x$, but has a negligible influence at larger distances.

From our scaling ansatz one obtains, after setting $\lambda = x - \sqrt{|\widehat k|} u$ and $z = x - \sqrt{|\widehat k|} y$:
\be
\delta g(z) = \dashint {\rm d}\lambda \frac{\delta \rho(\lambda)}{z-x+x-\lambda} \approx \text{i} \zeta \dashint {\rm d}u \frac{F(u)}{u-y}+ O(\widehat k).
\ee
Inserting such a scaling form into the right hand side of Eq. (\ref{saddle_disc}) yields a result independent of $\widehat k$ in the scaling regime:
\be
- \text{i}\widehat k \partial_x \left[\frac{\delta g(z)-\delta g(x)}{z-x}\right] =  \partial_y \left[\frac{\Gamma(0)-\Gamma(y)}{y}\right] + O(\widehat k),
\ee
where
\be
\Gamma(y) := \dashint {\rm d}u \frac{F(u)}{u-y}.
\ee
Now, we re-write Eq. (\ref{saddle_disc}) in terms of $g(z)=g_0(z) + \delta g(z)$, where $g_0(z)$ is the unperturbed Stieltjes transform, solution of:
\be 
g_0^2(z) - V'(z) g_0(z) + P(z) = 0 \Rightarrow g_0(z) = \frac12\left[V'(z) \pm \sqrt{V'^2(z) - 4P(z)}\right],
\ee
corresponding to an unperturbed eigenvalue density
\be
\rho_0(\lambda) = \frac{1}{2\pi} \sqrt{4P(\lambda) - V'^2(\lambda)}.
\ee
Neglecting terms of order $\widehat k$, the equation for $\delta g(z)$ then reads:
\be
 \delta g^2(z) + \left(2g_0(z) - V'(z)\right) \delta g(z)  =  2\partial_y \left[\frac{\Gamma(0)-\Gamma(y)}{y}\right]
\ee
or, for $z=x-\sqrt{|\widehat k|}y + \text{i} \zeta 0^+$ and $x$ inside the eigenvalue spectrum,
\be\label{ode}
- \frac12 \Gamma^2(y) + \pi \rho_0(x) \Gamma(y) =  \partial_y \left[\frac{\Gamma(0)-\Gamma(y)}{y}\right]
\ee
One can immediately deduces from this ODE that the asymptotic behaviour of $\Gamma(y)$ is:
\be 
\Gamma(y) \sim_{|y| \to \infty} \, - \frac{\Gamma(0)}{\pi \rho_0(x)} y^{-2} + O(y^{-4}),
\ee
or 
\be
\delta g(z) \sim_{|z-x| \gg \sqrt{|\widehat k|}} \, \, - \text{i} \widehat k \frac{\Gamma(0)}{\pi \rho_0(x)} (x-z)^{-2}.
\ee
Identifying with the asymptotic result Eq. (\ref{asympt}) garnered from the continuum approximation, we get an equation fixing the value of $\Gamma(0)$:
\be
\Gamma(0) =  \pi \rho_0(x).
\ee
Note that this result is super-universal, in the sense that it does not depend on $\beta$, nor on the shape of the confining potential $V(\lambda)$. Noting that Eq. (\ref{ode}) is of the Ricatti type and that 
$\Gamma(y)$ is a regular, even function of $y$, one finds that the solution can be expressed in terms of the modified Bessel function of the first type as: 
\be
\frac{\Gamma(y)}{\Gamma(0)} = 1 - \frac{\Psi'(v)}{\Psi(v)}; \qquad \text{with} \quad \Psi(v)= v^{3/4}I_{-\frac34}(v) \quad \text{and} \quad v:=\Gamma(0) \frac{y^2}{4},
\ee
which is plotted in Fig. 1 together with the corresponding density perturbation $F(u)$. The continuum limit completely disregards the non-monotonic nature of the function, while only retaining the $-y^{-2}$ behaviour for large arguments.

\begin{figure} 
\begin{center}
\includegraphics[scale=0.6]{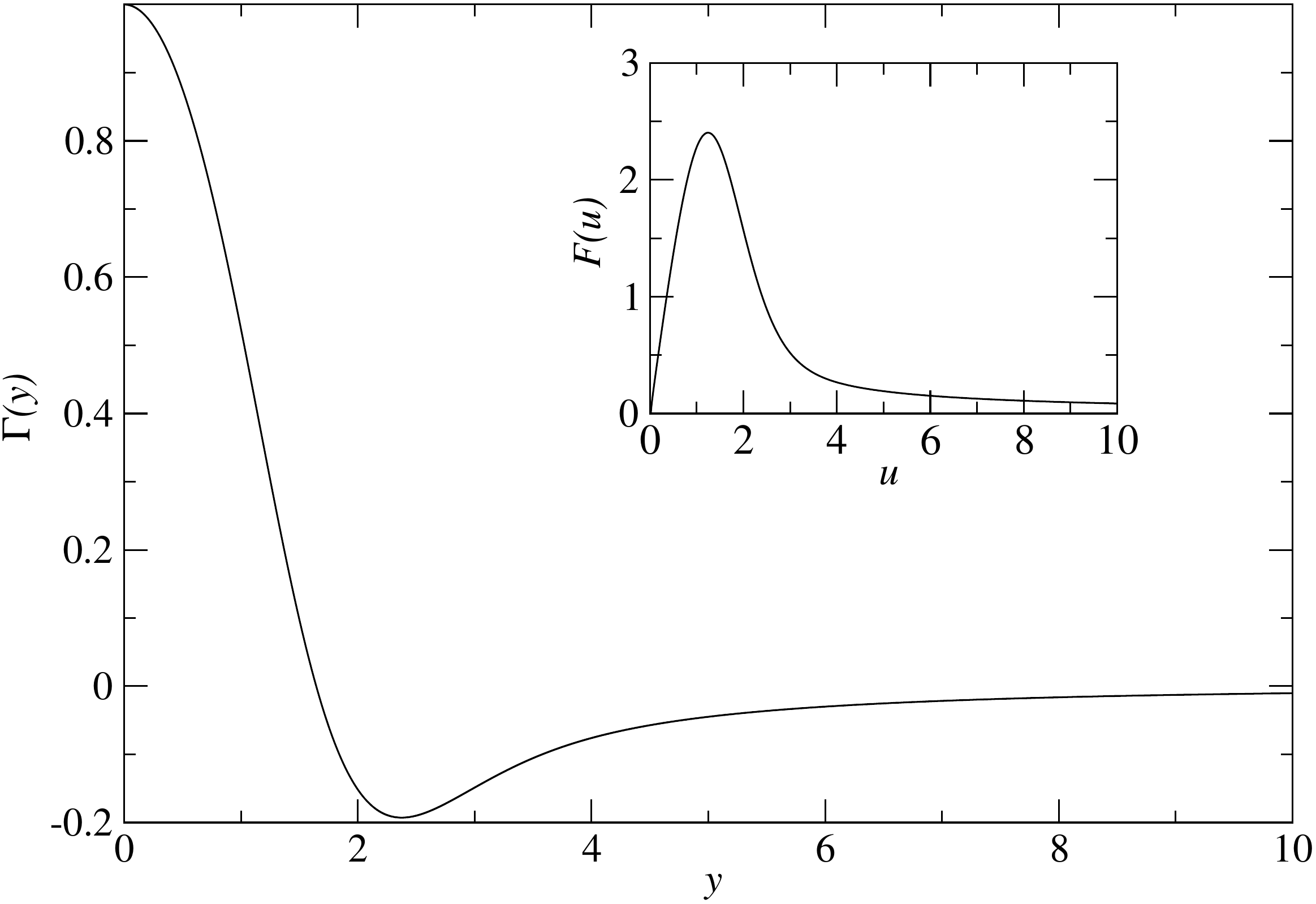}
\caption{Plot of the scaling function $\Gamma(y)$ for $y \geq 0$ and $\Gamma(0)=\pi \rho_0(x)=1$. Note the asymptotic behaviours $\Gamma(y) \approx 1 - y^2/2$ for small $y$ and $\approx -y^{-2}$ for large $y$.
Inset: Corresponding density perturbation scaling function $F(u)$, defined in Eq. (\ref{eq-defF}). Note that $F(-u)=-F(u)$.}
\label{fig_pnls}
\end{center}
\end{figure}

Endowed with these results, let us go back to our initial problem, which was to estimate the characteristic function of the distribution of $g$, Eq. (\ref{charac}). Using the 
fact that $\rho_0(\lambda)$ is a saddle point for $k=0$, one finds:
\be
\log \widehat P(k) \approx \text{i}k \int {\rm d}\lambda \frac{\rho_0(\lambda) + \delta \rho(\lambda)}{x - \lambda}.
\ee
The first contribution is simply the usual real part of the Stieltjes function $g_0(z)$ for $z=x$, given by $V'(x)/2$ when $x$ is inside the spectrum. The second term is imaginary and 
precisely given, to lowest order in $\widehat k$, by $\text{i} \zeta \Gamma(0)$. Therefore, for large $N$ (and thus small $\widehat k$) one finally obtains:
\be\label{final-result}
\log \widehat P(k) = \frac12\text{i}k V'(x) - |k| \pi \rho(x),
\ee
which corresponds precisely the Cauchy distribution obtained in the previous section, that decays for large $g$ as $\rho(x)/g^2$, as expected from general arguments. Our detailed description of the 
saddle-point solution may allow one to characterize finite $N$ and/or large deviation effects (see e.g. \cite{Grabsch}). 

\subsection{The L\'evy case}

Most of the arguments of section \ref{2.1} only rely on the fact that the elements of $\bf W$ are independent, identically distributed random random variables with a finite second moment, but not necessarily Gaussian. If the second moment is finite, we expect that all the above results will generalize, since they only rely on local properties of the spectrum (see \cite{Erdos, Aizenman}). Let us remind the reader how
such arguments must be adapted to the case where the matrix elements of $W$ have a diverging second moment, or more precisely when:
\be
P(W_{0i}) \sim_{W_{0i} \to \pm \infty} \frac{1}{N |W_{0i}|^{1+\mu}},
\ee
with $\mu < 2$. Then it is well known that the sum $\sum_{i}^N W_{0i}^2 G_{ii}$ converges towards a L\'evy stable distribution $L_{\mu/2}^{C,\beta}$ of index $\mu/2$, 
and scale and assymetry parameters respectively given by \cite{PhysRep}:
\be
C := \frac1N \sum_i |G_{ii}|^{\mu/2}; \qquad \beta:= \frac{1}{NC} \sum_i \text{sign}(G_{ii}) |G_{ii}|^{\mu/2}.
\ee
Since $S := x - 1/G_{00}$ is distributed according to $L_{\mu/2}^{C,\beta}(S)$, one deduces from Eq. (\ref{eq-iter}) that $G_{00}$ itself is distributed according to:
\be\label{PGL}
P(G_{00}) = \frac{1}{G_{00}^2} L_{\mu/2}^{C,\beta}(x-1/G_{00}).
\ee
Now, assuming that for large $N$ $G_{00}^{(N+1)}$ has the same distribution as the $G_{ii}^{(N)}$ allows one to find the following self-consistent relations for $C$ and $\beta$ for a given
value of $x$ \cite{cizeau,Burda,BAG}:
\bea 
C &=& \int_{-\infty}^{+\infty} {\rm d}G |G|^{\mu/2 - 2} L_{\mu/2}^{C,\beta}(x-1/G),\\  
C\beta &=&  \int_{-\infty}^{+\infty} {\rm d}G \, \text{sign}(G) |G|^{\mu/2 - 2} L_{\mu/2}^{C,\beta}(x-1/G).
\eea
Finally, the distribution of eigenvalues $\rho_{L}(x)$ of L\'evy matrices is obtained, as above, as the coefficient of the $G^{-2}$ tail of $P(G)$, i.e., from Eq. (\ref{PGL}):
\be
\rho_{L}(x) = L_{\mu/2}^{C,\beta}(x).
\ee
As shown in \cite{BAG}, this result coincides (rather miraculously) with the one obtained using the standard route, i.e. working in the complex plane, with $z = x - \text{i}\varepsilon$.
However, note that the distribution of a single diagonal element (say $G_{00}$) is no longer a Cauchy distribution (see Eq. (\ref{PGL})), although it shares the same power-law tail. 
The difference with the Wigner case lies in the fact that the L\'evy ensemble is not rotationally invariant, and is in fact characterized by strong correlations between eigenvectors and eigenvalues. In this case, there
is no reason to expect that the distribution of the normalized trace of $\G$ and of its diagonal elements is the same. Still, one knows from Aizenman \& Warzel that the super-universal Cauchy distribution also holds for the Stieltjes transform (or normalized trace) of L\'evy matrices. 

\section{A Dyson Brownian Motion for the Resolvant}

Since the seminal paper of Dyson in 1962 \cite{dyson1962brownian}, it is well known that the spectrum of Gaussian random matrices can be described in terms of the (fictitious) motion of $N$ interacting ``particles'' representing the position of the eigenvalues. More precisely, let us introduce a fictitious time $t$ and define a symmetric random matrix $\M(t)$ as:
\begin{equation}
	\label{eq:dGOE_BM}
	\M(t) = \C + \BBB(t)
\end{equation}
where the $W_{ij}(t)$, $i \leq j$ are independent and identically distributed real Brownian motions, of variance $ \sigma^2 t/N$ for $i\neq j$ and $2\sigma^2 t/N$ for $i= j$. As is well known, the dynamics of the eigenvalues of $\M(t)$ is then characterized by a stochastic differential equation (SDE), known as \emph{Dyson's Brownian motion}:
\begin{eqnarray}
	\label{eq:DBM_eigenvalues}
	\dd\lambda_i(t) & = & \sqrt{\frac{2 \sigma^2}{N}} \dd b_{i}(t) + \frac1N \sum_{j\neq i}^{N} \frac{\dd t}{\lambda_i(t) - \lambda_j(t)}, \nonumber \\
	\lambda_i(0) & = & \mu_i, 
\end{eqnarray}
for $i = 1,\dots,N$, and where the $b_i(t)$ are independent real Brownian motions. The initial conditions $\lambda_i(t=0)$ are given by the eigenvalues of $\C$, $\mu_1 \geq \mu_2 \geq \dots \mu_N$.

Here, we present an approach that considers directly the time evolution of the full resolvent matrix $\G(z,t)$, which we have not seen in the literature before its publication in our review paper \cite{PhysRep2}. To that end, we define
\begin{equation}
	\label{eq:resolvent_BM}
	\G(z,t) \;\deq\; (z\mathbb{I} - \M(t))^{-1}.
\end{equation}
Using It{\^o} formula and the fact that $\dd M_{kl} = \dd W_{kl}$, one has
\begin{eqnarray}
	\label{eq:G_DBM_step1}
	\dd G_{ij}(z,t)  & = &  \sum_{k,l=1}^{N} \frac{\partial G_{ij}}{\partial M_{kl}} \dd W_{kl} + \frac12 \sum_{k, l,m,n=1}^{N} \frac{\partial^2 G_{ij}}{\partial M_{kl} \partial M_{mn}} \dd \qb{W_{kl} W_{mn}},
\end{eqnarray}
where we have treated $M_{kl}$ and $M_{lk}$ as independent variables following 100\% correlated Brownian motions.
Next, we compute the derivatives:
\begin{equation}
	\label{eq:resolvent_derivatives_1}
	\frac{\partial G_{ij}}{\partial M_{kl}} = \frac12 \left[ G_{ik} G_{jl}+G_{jk} G_{il} \right],  
\end{equation}
from which we deduce the second derivatives
\begin{equation}
	\label{eq:resolvent_derivatives_2}
	\frac{\partial^2 G_{ij}}{\partial M_{kl} \partial M_{mn}} = \frac14 \left[ \left(G_{im} G_{kn} + G_{im} G_{kn} \right) G_{jl} + ... \right],
\end{equation}
where we have not written the other 6 $GGG$ products. Now, using the properties of the Brownian noise $\BBB$, the quadratic co-variation reads
\begin{equation}
	\label{eq:quadratic_covar}
	\dd\qb{W_{kl} W_{mn}} =  \frac{\sigma^2\dd t}{N} \pBB{ \delta_{km}\delta_{ln}+\delta_{kn}\delta_{lm} }
\end{equation}
so that we get from \eqref{eq:G_DBM_step1} and taking into account symmetries:
\begin{equation}
	\label{eq:G_DBM_step3}
	\dd G_{ij}(z,t)  =  \sum_{k,l=1}^{N} G_{ik} G_{jl} \dd W_{kl} + \frac{\sigma^2}{N} \sum_{k,l=1}^{N} \pB{G_{ik} G_{lk} G_{lj} + G_{ik} G_{kj} G_{ll}} \dd t\,.
\end{equation}
If we now take the average over with respect to the Brownian motion $W_{kl}$, we find the following evolution for the average resolvent:
\begin{equation}
	\label{eq:G_DBM_avg}
	\partial_t \mathbb{E} [\b G(z,t)] \;=\; \sigma^2 \stj(z,t) \, \mathbb{E} [\G^2(z,t)] +  \frac{1}{N} \mathbb{E} [\b G^{3}(z,t)].
\end{equation}
Now, one can notice that:
\begin{equation}
\G^2(z,t) = - \partial_z \G(z,t); \qquad \G^3(z,t) =  \frac12\partial^2_{zz} \G(z,t),
\end{equation}
which hold even before averaging. By sending $N\to\infty$, we obtain the following matrix PDE for the resolvent:
\begin{equation}
	\label{eq:G_DBM_avg_asymp}
	\partial_t \mathbb{E} [\b G(z,t)] \;=\; - \sigma^2 \stj(z,t) \, \partial_z \mathbb{E} [\b G(z,t)] \,, \quad\text{with}\quad 
	\mathbb{E} [\G(z,0)] \; = \; \G_\C(z)\,,
\end{equation}
Note that this equation is {\it linear} in $\b G(z,t)$ once the Stieltjes transform $\stj(z,t)$ is known.  
Taking the trace of Eq. (\ref{eq:G_DBM_avg_asymp}) immediately leads to a Burgers equation for $g(z,t)$ itself \cite{rodgersshi,allez2013eigenvectors}:
\begin{equation}
	\label{eq:St_DBM_avg_asymp}
	\partial_t \stj(z,t) \;=\; - \sigma^2 \stj(z,t) \, \partial_z \stj(z,t) \,, \quad\text{with}\quad 
	\stj(z,0) \; = \; \stj_\C(z)\, := \frac1N \Tr \, (z\mathbb{I} - \C)^{-1}.
\end{equation}
Its solution can be found using the method of characteristics and reads:
\begin{equation}
	\label{eq:dGOE_resolvent_DBM}
	\stj(z,t) = \stj_\C(Z(z,t)), \qquad Z(z,t)\;\deq\; z - \sigma^2 t \stj(z,t).
\end{equation}
It is plain to see that when $\C = 0$, one has $\stj_\C(Z)=1/Z$, leading to the familiar second degree equation for $g(z):=g(z,1)$:
\be
g(z) (z - \sigma^2 g(z)) = 1,
\ee
identical to Eq. (\ref{gz}). More generally, Eq. (\ref{eq:dGOE_resolvent_DBM}) is identical to the well known free addition law for R-transforms \cite{Verdu, PhysRep2}.

More interesting is the solution of Eq. \eqref{eq:G_DBM_avg_asymp} for the full resolvent, that simply reads \cite{shlyakhtenko1996random,allez2014eigenvectors}:
\begin{equation}
	\label{bc:dGOE_resolvent_DBM}
	\G(z,t) = \G_\C(Z(z,t)),
\end{equation}
as can be checked by inserting this ansatz in Eq. \eqref{eq:G_DBM_avg_asymp}, and making use of Eq. (\ref{eq:dGOE_resolvent_DBM}). One can use this result to
extract the mean squared overlap between the eigenvectors $\b u_{i}(t)$ of the perturbed matrix $\M$ and the unperturbed eigenvectors $\b u_{j}(0) = \b v_j$ of $\C$. 
Indeed, let us consider the following projection $\scalar{\b v_j}{G_{ii}(z,t) \b v_j}$ with $z = \lambda_i -\text{i} \varepsilon$. In the large $N$ limit, this 
quantity converges to
\begin{equation}
  \scalar{\b v_j} {{\b G}(z,t) {\b v_j}} \underset{N \uparrow \infty}{\sim} \int \frac{\mso(\lambda, \mu_j) \rho_{\M}(\lambda) }{\lambda_i - \lambda - \ii \varepsilon} \dd \lambda,
\end{equation}
where $\varepsilon \gg N^{-1}$ and $\mso(\lambda, \mu_j)$ is the smoothed squared overlap between the eigenvector of $\C$ associated with eigenvalue $\mu_j$ and  
eigenvectors of $\M$ around eigenvalue $\lambda$, averaged over a small interval of width $\varepsilon$. Therefore, one gets
\begin{equation}
  \label{eq:resolvent_overlap}
  \mso(\lambda_i, \mu_j) = \frac{1}{\pi\rho_{\M}(\lambda_i)} \lim_{\varepsilon \to 0^{+}} \Im \scalar{\b v_j}{ \b G(\lambda_i - \ii \varepsilon,t) \b v_j}.
\end{equation}
Using Eq. (\ref{bc:dGOE_resolvent_DBM}) with $\scalar{\b v_j}{\G_\C(Z) \b v_j} = (Z - \mu_j)^{-1}$, one finally obtains, for $t=1$: 
\begin{equation}
	\label{eq:overlap_DBM}
	N \mathbb{E} \qb{\scalar{\b u_i}{\b v_j}^2} = \frac{\sigma^2}{(\lambda_i - \mu_j - \sigma^2 \Re\stj(\lambda_i))^2 + \sigma^4 \pi^2 \rho^2(\lambda_i)}.
\end{equation}
This result was first obtained in \cite{shlyakhtenko1996random,allez2013eigenvectors}, and is the counterpart of the Ledoit and P\'ech\'e result \cite{ledoit2011eigenvectors} in the 
context of multiplicative models (see \cite{IEEE, PhysRep2} for more details).  Note that the square overlap is of order $N^{-1}$ as soon as $\sigma > 0$. For a given $\lambda_i$, 
the overlap has a Lorentzian shape as a function of $\mu_j$, that peaks at $\lambda_i - \sigma^2 \Re\stj(\lambda_i)$. 

Note that Eq. (\ref{eq:dGOE_resolvent_DBM}) obviously generalizes to any intermediate time $t_0$ as
\begin{equation}
	\label{eq:dGOE_resolvent_DBM_bis}
	\stj(z,t) = \stj(Z(z,t;t_0),t_0), \qquad Z(z,t;t_0)\;\deq\; z - \sigma^2 (t-t_0) \stj(z,t),
\end{equation}
and correspondingly
\begin{equation}
	\label{bc:dGOE_resolvent_DBM_bis}
	\G(z,t) = \G(Z(z,t;t_0),t_0).
\end{equation}
This enables one to compute the eigenvector overlaps between any two times $t_0$ and $t$. 

Another interesting case is when the initial matrix $\C$ is of rank one, with a single non zero eigenvalue $\mu_1$ and eigenvector $\b v_1$, and all the other $N-1$ ones are zero, as above. 
If one carefully keeps terms of order $1/N$, the whole formalism 
allows one to keep track of the isolated eigenvalue for $t > 0$, and the corresponding overlap $\Phi_1:=\scalar{\b u_1}{\b v_1}^2$. One readily finds that the isolated eigenvalues persists up to $t = t^* = (\mu_1/\sigma)^2$, 
and is located at \cite{Peche}:
\be 
\lambda_1(t) = \mu_1 + \frac{\sigma^2 t}{\mu_1},
\ee
before colliding with the edge of the Wigner spectrum $2 \sigma \sqrt{t}$ precisely at $t=t^*$ and disappearing altogether in the Wigner sea for $t > t^*$. This is the famous BBP transition \cite{BBAP}. The overlap $\Phi_1$ is contained in the corresponding pole of $\G(z=\lambda_1,t)$, as given by Eq. (\ref{eq:dGOE_resolvent_DBM}) and is found to be given by:
\be
\Phi_1(t) = 1 - \frac{t}{t^*},
\ee
which goes to zero at $t=t^*$, as it should be \cite{BBP,allez2014eigenvectors}. The way $\Phi_1(t^*)$ behaves for finite $N$ is apparently not known. A natural conjecture is that, for $t \approx t^*$ and $N \gg 1$, the following scaling result holds:
\be
\Phi_1(t,N) = N^{-1/3} \varphi\left(N^{1/3}\frac{t^*-t}{t^*}\right),
\ee
where $\varphi(u \gg 1)=u$ and $\varphi(u=0)=\varphi_0$ a positive constant. 

Finally, it is interesting to write the dynamical equation for $\stj(z,t)$ keeping terms of order $1/N$, since some non zero noise survives in that limit. One finds the following Langevin equation:
\be
\partial_t \stj(z,t) \;=\; - \sigma^2 \stj(z,t) \, \partial_z \stj(z,t) + \frac1{2N} \partial_{zz} \stj(z,t) + \frac{\sigma}{N} \xi(z,t),
\ee
where $\xi$ is a white (Langevin) complex noise, such that:
\be
\mathbb{E}(\xi(z,t) \xi(z,t')) = - 2 \partial_{zzz} \stj(z,t) \delta(t-t').
\ee
It would be interesting to see if one can extract some useful information from this formalism. In particular, is it possible to recover the Cauchy distribution discussed in the first part of this paper as 
a stationary distribution of the above Langevin equation?

It would also be interesting to generalize the Dyson approach to the resolvent of covariance matrices. Work in that direction is underway \cite{inprep}. 

\vskip 1cm

We thank R. Allez, G. Biroli, J. Bun, Y. Fyodorov, A. Guionnet,  V. Hakim and D. Shlyakhtenko for very fruitful exchanges on these topics. 
\newpage

\appendix
\section{The Aizenman-Warzel approach} 

The paper of Aizenman \& Warzel \cite{Aizenman} is not easy to penetrate (at least for us). Here we give a simplified version of their approach, that allows one to understand the super-universal nature of the Cauchy law. Assume that instead of fixing the (real) value of $x$ at which we want to compute the distribution of Stieltjes transform $g(x)$ over the considered ensemble of random matrices, one rather fixes the position of the eigenvalues $\l_i$ in a typical configuration, such that the local density is $\rho(x)$ and the typical distance between consecutive eigenvalues is $(N \rho(x))^{-1}$. We now want to study the distribution of $g(x+ \eta u)$, with $u \sim O(1)$ a random variable with an arbitrary distribution $R(u)$ and $\eta$ a small parameter such that $N^{-1} \ll \eta \ll 1$. The characteristic function of the distribution of $g$ is then:
\be \label{charac2}
\widehat P(k) = \int \ {\rm d}u \, R(u) \, e^{\text{i}{k} g(x + \eta u)}.
\ee
One of the lemma of \cite{Aizenman} is that the choice of $R(u)$ is arbitrary provided the range of $\eta u$ covers many eigenvalues, i.e. $N^{-1} \ll \eta$. They choose 
\be
R(u)=\frac{1}{\pi} \frac{1}{1+u^2},
\ee
i.e. a Cauchy distribution -- but unrelated to the final (Cauchy) result we are looking for!\footnote{One can easily check that taking other forms that allow one to use residues, such as $R(u) =(2/\sqrt{3}\pi) (1+u^2)^{-2}$, leads to the same final result.} Now, one should note that the simple pole structure of $g(x)$ implies that it maps the upper complex plane $\mathbb{C}^+$ into the lower complex plane $\mathbb{C}^-$. Hence, $\exp[\text{i}{k} g(x + \eta u)]$ is bounded when $u$ is in the lower (resp. upper) complex plane when $k > 0$ (resp. $k < 0$). For $k> 0$, one can therefore calculate Eq. (\ref{charac2}) using a contour integration in the lower complex plane, enclosing the pole of $R(u)$ at $u=-\text{i}$. The result is: 
\be 
\widehat P(k) =  e^{\text{i}{k} g(x - \text{i} \eta)}.
\ee
Similarly, for $k < 0$, the pole is at $u=+\text{i}$ and:
\be 
\widehat P(k) =  e^{\text{i}{k} g(x + \text{i} \eta)}.
\ee
For large $N$, $g(x+\text{i} \eta)$ converges to the limiting Stieltjes distribution $g_0(x+\text{i} \eta)$ provided that $\eta\gg 
N^{-1}$.  Taking the limit $\eta \to 0$ in this sense leads to the characteristic function of the Cauchy distribution, identical to Eq. (\ref{final-result}):
\be 
\widehat P(k) =  e^{\text{i}{k} g_R(x)  - |k|\pi \rho(x)},
\ee
where we have used the standard result:
\be
\lim_{\eta \to 0^+} g_0(x - \text{i} \eta) = g_R(x) - \text{i} \pi \rho(x).
\ee
That the process of averaging over the matrix ensemble or over the position of $x$ leads to the same result is not unexpected, but not totally trivial either. 

\section{The case of a periodic array of eigenvalues} 

Assume that the eigenvalues are locally equally spaced. Near a certain $x$, eigenvalues are spaced by $\Delta=1/(N\rho(x))$. $\rho(x)$ is assumed to be constant for $L=\sqrt{N}$ eigenvalues above and below $\lambda_m$, defined as the closest eigenvalue to $x$.

Let $x-\lambda_m=\Delta u$, so $u$ is uniform in $[-1/2,1/2]$.
For the $L$ first eigenvalues larger than $\lambda_m$, we have $x-\lambda_{m+k}=\Delta(u+k)$. For the $L$ eigenvalues immediately below $\lambda_m$ we have $x-\lambda_{m-k}=\Delta(u-k)$. 
We split $g(x)$ into two parts, one near $x$ one far from $x$:
\be
g(x)=\frac{1}{N}\left[
\sum_{k\notin[m-L,m+L]}\frac{1}{x-\lambda_k}
+\frac{1}{\Delta u}
+\sum_{k=1}^{L}\left(\frac{1}{\Delta(u-k)}+\frac{1}{\Delta(u+k)}\right)
\right]
\ee
The first sum can be replaced by a principal part integral, using $\rho(x)=1/(N\Delta)$ and grouping the last two terms we get
\be\label{g_flat}
g(x)= g_R(x) + \rho(x)\left(\frac{1}{u}+2u\sum_{k=1}^{L}\frac{1}{u^2-k^2}\right),
\ee
where $g_R(x)$ is again the real part of the limiting Stieltjes transform. The sum on the right is convergent, we can replace $L=\sqrt{N}\to\infty$. Mathematica says this sum is
\be
\frac{1}{2u}\left(\pi\cot(\pi u)-\frac{1}{u}\right)
\ee
We can rewrite Eq.\ (\ref{g_flat}) as
\be
\frac{1}{\pi\rho(x)}\left(g(x)-g_R(x)\right)=\cot(\pi u)
\ee
\be
u=\pi^{-1}\cot^{-1}\left(\frac{g(x)-g_R(x)}{\pi\rho(x)}\right)
\ee
which is equivalent to saying that $g(x)$ is distributed according to the Cauchy law centered at $g_R(x)$ and of width $\pi\rho(x)$, as for the Poisson case, or any other value of $\beta$ for Coulomb gas models.


\begin{thebibliography}{99}

\bibitem{Fyodorov1} Fyodorov, Y. V., \& Sommers, H. J. (1997). Statistics of resonance poles, phase shifts and time delays in quantum chaotic scattering: Random matrix approach for systems with broken time-reversal invariance. Journal of Mathematical Physics, 38(4), 1918-1981.

\bibitem{Fyodorov2} Fyodorov, Y. V., \& Savin, D. V. (2004). Statistics of impedance, local density of states, and reflection in quantum chaotic systems with absorption. Journal of Experimental and Theoretical Physics Letters, 80(12), 725-729.

\bibitem{Fyodorov3} Fyodorov, Y. V., \& Williams, I. (2007). Replica symmetry breaking condition exposed by random matrix calculation of landscape complexity. Journal of Statistical Physics, 129(5-6), 1081-1116.

\bibitem{Aizenman} Aizenman, M., \& Warzel, S. (2015). On the ubiquity of the Cauchy distribution in spectral problems. Probability Theory and Related Fields, 163(1-2), 61-87.

\bibitem{cizeau} Cizeau, P., \& Bouchaud, J. P. (1994). Theory of L\'evy matrices. Physical Review E, 50(3), 1810.

\bibitem{Burda} Burda, Z., Jurkiewicz, J., Nowak, M. A., Papp, G., \& Zahed, I. (2007). Free random L\'evy and Wigner-L\'evy matrices. Physical Review E, 75(5), 051126.

\bibitem{BAG} Ben Arous, G., \& Guionnet, A. (2008). The spectrum of heavy tailed random matrices. Communications in Mathematical Physics, 278(3), 715-751.

\bibitem{Hakim} Griniasty, M., \& Hakim, V. (1994). Correlations and dynamics in ensembles of maps: Simple models. Physical Review E, 49(4), 2661.

\bibitem{Majumdar} Dean, D. S., \& Majumdar, S. N. (2008). Extreme value statistics of eigenvalues of Gaussian random matrices. Physical Review E, 77(4), 041108.

\bibitem{Erdos} Erdos, L., Schlein, B., \& Yau, H. T. (2010). Wegner estimate and level repulsion for Wigner random matrices. International Mathematics Research Notices, 2010(3), 436-479.

\bibitem{Grabsch} Grabsch, A., \& Texier, C. (2016). Distribution of spectral linear statistics on random matrices beyond the large deviation function : Wigner time delay in multichannel disordered wires. Journal of Physics A: Mathematical and Theoretical, 49(46), 465002.

\bibitem{PhysRep} Bouchaud, J. P., \& Georges, A. (1990). Anomalous diffusion in disordered media: statistical mechanisms, models and physical applications. Physics reports, 195(4-5), 127-293.

\bibitem{dyson1962brownian} Dyson, F. J. (1962). A Brownian-motion model for the eigenvalues of a random matrix. Journal of Mathematical Physics, 3(6), 1191-1198.

\bibitem{PhysRep2} Bun, J., Bouchaud, J. P., \& Potters, M. (2017). Cleaning large correlation matrices: tools from random matrix theory. Physics Reports, 666, 1-109.

\bibitem{rodgersshi} Rogers, L. C. G., \& Shi, Z. (1993). Interacting Brownian particles and the Wigner law. Probability theory and related fields, 95(4), 555-570.

\bibitem{allez2013eigenvectors} Allez, R., \& Bouchaud, J. P. (2014). Eigenvector dynamics under free addition. Random Matrices: Theory and Applications, 3(03), 1450010.

\bibitem{allez2014eigenvectors} Allez, R., Bun, J., \& Bouchaud, J. P. (2014). The eigenvectors of gaussian matrices with an external source. arXiv preprint arXiv:1412.7108.

\bibitem{Verdu} see e.g. Tulino, A. M., \& Verd\'u, S. (2004). Random matrix theory and wireless communications. Foundations and Trends in Communications and Information Theory, 1(1), 1-182.

\bibitem{shlyakhtenko1996random} Shlyakhtenko, D. (1996). Random Gaussian band matrices and freeness with amalgamation. International Mathematics Research Notices, 1996(20), 1013-1025.

\bibitem{ledoit2011eigenvectors} Ledoit, O., \& P\'ech\'e, S. (2011). Eigenvectors of some large sample covariance matrix ensembles. Probability Theory and Related Fields, 151(1-2), 233-264.

\bibitem{IEEE} Bun, J., Allez, R., Bouchaud, J. P., \& Potters, M. (2016). Rotational Invariant Estimator for General Noisy Matrices. IEEE Trans. Information Theory, 62(12), 7475-7490.

\bibitem{Peche} F\'eral, D., \& P\'ech\'e, S. (2007). The largest eigenvalue of rank one deformation of large Wigner matrices. Communications in mathematical physics, 272(1), 185-228.

\bibitem{BBAP} Baik, J., Ben Arous, G., \& P\'ech\'e, S. (2005). Phase transition of the largest eigenvalue for non-null complex sample covariance matrices. The Annals of Probability, 33(5), 1643-1697.

\bibitem{BBP} Biroli, G., Bouchaud, J. P., \& Potters, M. (2007). On the top eigenvalue of heavy-tailed random matrices. EPL (Europhysics Letters), 78(1), 10001.

\bibitem{inprep} Bun, J., Bouchaud, J. P., \& Potters, M. In preparation.

\end{thebibliography}
\end{document}